\documentclass[fleqn,usenatbib]{mnras}

\usepackage{newtxtext,newtxmath}

\usepackage[T1]{fontenc}

\DeclareRobustCommand{\VAN}[3]{#2}
\let\VANthebibliography\thebibliography
\def\thebibliography{\DeclareRobustCommand{\VAN}[3]{##3}\VANthebibliography}

\usepackage{graphicx}	
\usepackage{amsmath}	
\usepackage[svgnames]{xcolor}
\usepackage{caption}
\usepackage{subcaption}

\title[Probing down to early cosmic epochs using limited redshift ($z\lesssim1$)]{Probing down to early cosmic epochs using limited redshift ($z\lesssim1$) optical surveys with host galaxy age and lookback time analysis}

\author[S. Kasthurirangan et al.]{
S. Kasthurirangan,$^{1}$\thanks{E-mail: s.kasthurirangan@physics.mu.ac.in (SK)}
M. Panchal$^{1}$,
A. Joshi$^{1}$,
G. Pawar$^{2}$
\\
$^{1}$Department of Physics, University of Mumbai, Vidyanagari, Mumbai - 400098, India\\
$^{2}$Nicolaus Copernicus Astronomical Center, Polish Academy of Sciences, ul. Rabiańska 8, 87-100 Toruń, Poland\\
}

\date{Accepted XXX. Received YYY; in original form ZZZ}

\pubyear{2024}

\begin{document}
\label{firstpage}
\pagerange{\pageref{firstpage}--\pageref{lastpage}}
\maketitle

\begin{abstract}
We present a detailed analysis of AGN identification diagnostics and host galaxy evolution using optical spectral diagnostics using ESO-GOODS-S data. We employ traditional Baldwin-Phillips-Terlevich (BPT) diagrams along with their modern extensions, the Mass-Excitation (MEx) and Colour-Excitation (CEx) diagrams, to classify AGNs from among up to 600+ candidates. We extract the spectral properties utilising an indigenous spectral fitting code. In addition, the code also incorporates inputs from a state-of-the-art stellar population synthesis model (Bruzual and Charlot, 2003, MNRAS, 344, 1000), extracting the host galaxy ages and metallicities as model parameters. Redshift values for all the studied spectra are obtained from the literature. Using these redshift values, we are able to assign a `lookback time' (using standard $\Lambda$CDM cosmological parameters). When this lookback time is combined with the galaxy ages (from spectral fitting), it enables us to study galaxy evolution even up to early cosmological epochs ($\sim$13 Gyr), despite restricting the studied field of galaxies to $z\sim1$ (lookback time of $\sim$8 Gyr). We conclude that galaxies ultimately evolving into AGN hosts originate earlier ($\sim8$ Gyr ago) than non-AGN host galaxies ($\sim6$ Gyr ago). We relate this fact, that AGN hosts date back to an earlier epoch than non-AGN galaxies, as being likely due to higher cosmic temperatures at that time, probably enhancing black hole accretion. We emphasise that host galaxy age is an additional crucial parameter (apart from cosmological redshift) in defining our understanding of AGN and star formation co-evolution across cosmological time scales.

\end{abstract}

\begin{keywords}
galaxies: active -- galaxies: evolution -- galaxies: formation -- galaxies: nuclei -- techniques: spectroscopic -- methods: data analysis
\end{keywords}



\section{Introduction \label{Intro}}

Active Galactic Nuclei (AGNs) are some of the most energetic and luminous systems in the universe, driven by the process of accretion onto supermassive black holes (SMBH) at the centers of galaxies \citep{1984_Rees,1995_Urry,1999_Krolik_Book}. The identification and characterisation of AGNs presents significant challenges, because no single observational indicator can fully capture the diverse and complex nature of black hole accretion and the accompanying wide-spectrum emission \citep{1993_Antonucci,2017_Padovani}. AGNs manifest with a wide range of observable signatures, depending on factors such as orientation, obscuration, and the surrounding environment~\citep{1995_Urry,2015_Netzer,2006_Hopkins}. Many AGNs exhibit varying degrees of activity across different wavelengths, and their evolutionary processes can be influenced by multiple factors, including interactions with their host galaxies, cosmic environment, and merger history \citep{2006_Hopkins,2012_Alexander,2017_Kalfountzou}.

Studies in different spectral wavebands probe distinct physical processes that contribute to the AGN phenomenon. For instance, optical emission lines reveal the ionized gas in the narrow- and broad-line regions \citep{1981_Baldwin_PASP}, while X-ray and UV wavelengths trace the highly energetic radiation from the innermost regions near the black hole \citep{Silverman_2005}. Infrared observations detect reprocessed radiation from dust heated by the AGN, especially in obscured systems \citep{2005_Stern}. Meanwhile, radio observations can detect relativistic jets and synchrotron emission, providing insights into the AGN's interaction with its host galaxy and surrounding medium~\citep{1984_Begelman}.

Optical and infrared diagnostics have been essential in differentiating AGNs from star-forming regions in galaxies. One of the most widely used methods is the classical Baldwin-Phillips-Terlevich (BPT) diagram \citep{1981_Baldwin_PASP}, which distinguishes AGNs from star-forming regions based on emission line ratios. This method is especially suited for dealing with wide field survey data, where the emphasis is on statistical trends rather than specific physical properties of the individual objects or regions under consideration. This diagram compares specific optical line ratios, such as [\ion{O}{III}]$\lambda$5007/H$\beta$ against [\ion{N}{II}]$\lambda$6584/H$\alpha$, to reveal the ionisation mechanisms at play within a galaxy. AGNs, which generate harder ionising radiation than star-forming regions \citep{1985_Osterbrock_Pogge_APJ}, occupy a distinct region of the BPT diagram, providing a seemingly clear diagnostic to identify their presence. More recent updates to the BPT method \citep{2006_Kewley, 2006_Stansiska} further refine these diagnostics, extending their applicability across a wider range of object types (e.g. Low Ionisation Nuclear Emission Line Regions (LINERs) vs Seyfert galaxies etc.). Studying the possibility of modifying this classification scheme across different redshift regimes has also been discussed in the literature~\citep{2013_Kewley, 2013_Trump, 2015_Coil}.

Beyond the BPT diagram, newer methods like the Mass-Excitation (MEx) and Color-Excitation (CEx) diagrams offer complementary approaches to differentiate AGNs from star-forming regions, particularly at low-to-intermediate redshifts ($z\lesssim 1$). The MEx diagram, introduced by \cite{2011_Juneau_APJ}, incorporates the stellar mass of the galaxy as the ordinate, while retaining the [\ion{O}{III}]$\lambda$5007/H$\beta$ line ratio as the abcissa. This enables the MEx diagram to capitalise on the observed correlation between mass and ionisation properties. Similarly, \cite{2011_Yan_APJ} studies the same line ratio against the (U-B) color index of the galaxy.  This approach is especially effective in identifying AGNs in dusty, obscured environments where optical diagnostics alone may fall short.



One of the aims of this paper is to present a detailed study of the Great Observatories Origins Deep Survey South (GOODS-S) field data using spectral analysis to better understand the properties of galaxies, with a particular focus on AGNs. In order to do that, we use an indigenously developed spectral code \citep{2024asi..confP..53P}, which addresses key spectral features, including the host galaxy contribution, AGN emission continuum, and broad and narrow spectral lines, etc. Using the spectral fitting parameters obtained from the code, we construct various AGN diagnostic diagrams, like traditional BPT diagrams as well as the modified ones, like the MEx and CEx. A second major aim of this paper is to introduce one of the key features of the code, which is the integration of template spectra from the ~\cite{2003_Bruzual_Charlot_MNRAS} stellar population synthesis model (hereafter, BC03) into our fitting process, enabling us to provide estimates of the ages of the host galaxies, as model fitting parameters. Introducing host galaxy age as an independent parameter takes the AGN diagnostics beyond its traditional paradigm, making it possible to study the temporal evolution of the sources over different cosmological epochs. Such a holistic approach could eventually lead to a better understanding of the role of AGNs in the evolution of the universe. Such approaches are seen in the existent literature \citep{2013_Kewley,2014_Goulding}, but none of these take into account the age of the host galaxy as an independent parameter. This is a unique feature of this study.

The paper is organised as follows. In \autoref{Data_section}, we describe the GOODS-S data used in our analysis. We also detail the selection criteria in terms of the availability of redshift data, and also in terms of redshift limits of the method used. In \autoref{Fitting Section}, we outline the spectral fitting process, including the integration of template spectra from the BC03 stellar population synthesis models into the same. The key flux ratios which form the basis for constructing traditional BPT diagnostic diagrams (\textit{e.g.}, [\ion{O}{III}/H$\beta$], [\ion{N}{II}/H$\alpha$], [\ion{S}{II}/H$\alpha$]), are obtained from the fitting parameters. In \autoref{Results and Discussion}, we present the results, including the classification of sources based on the traditional BPT diagrams ([\ion{N}{II}], [\ion{S}{II}], [\ion{O}{II}]; see~\autoref{BPTs_trad}) and more recent MEx and CEx diagrams (see~\autoref{MEx_CEx}). Further, the galaxy ages estimated from the host galaxy fitting is then used to study trends in normal and active galaxies (see~\autoref{Temporal Evolution}). This section also explores the correlation between the derived ages of the sources and the lookback time (based on redshift), offering insights into how AGNs and star-forming galaxies evolve over different epochs. The broader conclusions from our findings are discussed in \autoref{Conclusion}, which also outlines the future directions for this line of research.



\section{DATA}

\label{Data_section}

The GOODS-S field, centered at RA 3$^h$ 32$^m$ 32.74$^s$ and DEC -27$^\circ$ 47' 20.40", spans 7.47 arcminutes and is characterized by its low hydrogen column density, minimizing galactic reddening and enabling clear extragalactic observations \citep{2003_Giavalisco}. This field has been the focus of extensive multiwavelength surveys, starting with the \textit{Chandra Deep Field South (CDFS)}, which provides the deepest X-ray imaging to date, complemented by wider-field \textit{XMM-Newton} observations \citep{2017_Luo,2011_Xue}. UV data have been contributed by \textit{GALEX} and \textit{Hubble}’s UV Legacy Survey, probing young stellar populations and star formation in early galaxies \citep{2007_Martin,2018_Teplitz}. Optical observations include deep \textit{HST} imaging with the ACS and WFC3 instruments, as part of the GOODS program, and spectroscopic studies from ground-based facilities like the VLT and Subaru, offering morphological classifications and redshift measurements \citep{2004_Giavalisco,2020_Cooper}. Infrared observations from \textit{Spitzer} and \textit{Herschel} have uncovered obscured star formation and thermal dust emission, while \textit{JWST} now delivers unprecedented near- and mid-infrared data, resolving galaxies at \(z > 10\) \citep{2011_Lutz,2023_Finkelstein}. Radio observations with the VLA at 1.4 GHz and submillimeter data from ALMA and ATCA have provided insights into star formation rates, molecular gas reservoirs, and dust content in galaxies \citep{2008_Morrison,2016_Dunlop}. These multiwavelength datasets collectively make GOODS-S an unparalleled field for investigating galaxy formation, AGN activity, and the cosmic history of baryonic matter.

We have utilized optical spectral data from the ESO GOODS-S survey for our analysis. The data were obtained from the ESO data archive, observed using the FORS1 \citep{FORS1}, FORS2 \citep{FORS2}, and VIMOS \citep{VIMOS} instruments mounted on the VLT at Paranal Observatory in Chile. The data sample analyzed in this study includes 8864 spectra covering the optical and near-optical wavelength regions, spanning from approximately 3600~\AA{} to 10,000~\AA{} \citep{1998_Appenzellar}. One of the aims of this study is to evaluate the various optical line-ratio-based classification schemes for AGNs, hence we primarily focus on optical spectroscopy data.

In addition to this primary dataset obtained from ESO, the redshift values of many of these sources were obtained separately from various surveys available for the same field, which are listed in~\autoref{redshift_surveys} to complete our dataset. The spectra were mapped to the redshift surveys using a methodology available in the Python libraries such as \href{https://docs.astropy.org/en/stable/coordinates/matchsep.html}{Astropy and TOPCAT integrations} \citep{2022_Astropy}, which uses RA and DEC values to match the sources across different surveys. This mapping was implemented using a Python-based code. From the available surveys, we mapped redshifts for 3534 unique sources.

In order to classify AGNs and star-forming galaxies, we have utilized traditional BPT diagrams~\citep{1981_Baldwin_PASP}, such as the [\ion{S}{II}], [\ion{N}{II}], and [\ion{O}{II}] diagnostic diagrams, which are commonly used to classify AGNs and star-forming galaxies based on their emission line ratios. These diagrams provide a clear separation between different types of galaxies, including AGNs, \ion{H}{II} regions, and composite galaxies. The [\ion{S}{II}] $\lambda$$\lambda$6717,6731, [\ion{N}{II}] $\lambda$6584, and [\ion{O}{II}] $\lambda$3727 emission lines are sensitive to the ionization state of the gas, making them key tracers of ionization mechanisms. 

The flux ratios of [\ion{O}{III}] $\lambda$5007/H$\beta$, [\ion{N}{II}] $\lambda$6584/H$\alpha$, and [\ion{S}{II}] $\lambda$$\lambda$6717,6731/H$\alpha$ are used to distinguish between star-forming galaxies and AGNs, with the latter showing strong emission from the AGN ionizing source. This methodology follows the diagnostic criteria developed by \cite{1981_Baldwin_PASP}, \cite{1987_Veilleux}, and further refined by \cite{2006_Kewley} to interpret the gas excitation and ionization mechanisms across a wide range of redshifts and galaxy types.

We have also utilized the MEx diagram~\citep{2014_Juneau}, which uses galactic stellar mass as one of its axes. The stellar mass, an important parameter for understanding galaxy evolution, represents the mass present in stars in the galaxy. For this analysis, we have mapped the stellar mass data from the Cosmic Assembly Near-infrared Deep Extragalactic Legacy Survey (CANDELS) by \cite{2015_Santini_APJ} to the objects whose spectroscopy data is considered by us.

We have also used the CEx diagram in this study~\citep{2011_Yan_APJ}, which requires the ($U-B$) color index as one of the axes. In general, the color index measures the difference between two measurements of the magnitude of an object made at different bands, with the longer wavelength value subtracted from the shorter wavelength value. The lower the color index, the bluer (or hotter) the object is, while a larger color index indicates a redder (or cooler) object. We have mapped the (U-B) color index data from the COMBO-17 survey by \cite{2004_Wolf_AAP}. 

We have also mapped the optical spectra to X-ray sources from the CDFS by \cite{2017_Luo}. X-ray luminosity data is a crucial metric for determining obscured AGNs, as it provides a direct probe of high-energy processes often hidden from optical wavelengths by dust and gas \citep{2001_Gilli}. X-ray AGN sources can also serve as a confirmation for optically identified AGNs, particularly in cases where optical diagnostics are ambiguous or where sources are heavily obscured in the optical. Additionally, the X-ray classifications from \cite{2017_Luo} provide a means for identifying trends and relating them to our AGN classifications, allowing for cross-verification and improved characterization of the AGN population. These combined datasets enable robust analysis of AGN properties, such as their luminosity functions, obscuration levels, and host galaxy environments \citep{2014_Buchner}.

In the study of AGNs, the gas clouds surrounding the central engine are pivotal for understanding their spectral features \citep{2006_Schneider_Book}. These clouds, typically found in spiral galaxies hosting AGNs, absorb ultraviolet and X-ray emissions from the central engine, leading to the emission of characteristic spectral lines \citep{1985_Osterbrock_Pogge_APJ}. These lines include strong hydrogen (H) lines and emission lines like [\ion{N}{II}]$\lambda$6584 and [\ion{O}{III}]$\lambda$5007, similar to those found in \ion{H}{II} regions of our galaxy \citep{2006_Schneider_Book,2013_Kewley}. These spectral features provide insights into the composition and density of the gas clouds, categorized into two regions: the broad-line region (BLR) with denser and warmer clouds and the narrow-line region (NLR) with lower-density and colder clouds, both influenced by the AGN's strong gravitational field \citep{2015_Netzer,1999_Krolik_Book}. According to the AGN unified model \citep{1993_Antonucci_ARAA,2017_Padovani}, these two regions are not found in fundamentally different classes of AGNs but rather reflect the observer’s line of sight with respect to the dusty torus surrounding the nucleus. When the central engine and the BLR are viewed directly (``face-on'', type 1 AGN), both broad and narrow emission lines are observed. However, if the line of sight is obscured by the torus (``side-on'', type 2 AGN), only the narrow-line region is visible. This orientation-dependent framework explains the diversity in AGN spectra while reinforcing that BLR and NLR are intrinsic components of all AGNs, while their observability is dependent on the viewing geometry.

The process of selection of data finally used in our analysis is outlined in~\autoref{data_reduction}. Initially, we have 8864 ESO spectra (Stage 1). Reliable redshifts are available only for 4845 spectra (Stage 2, see~\autoref{redshift_surveys}). In stage 3, we map these 4845 spectra to 3534 unique sources. Of these, the data is of sufficient quality to undertake fitting with our model only in 2937 cases (Stage 4). These include even those with only host galaxy fitting. In the remaining cases, the fitting algorithm was failing due to poor S/N ratio. Further, spectra of only 726 unique sources are found to be in the wavelength range suitable for use in the diagnostic diagrams used in this study, since [\ion{O}{III}]$\lambda$5007/H$\beta$ ratio must be available for this (Stage 5). This corresponds to a minimum corrected wavelength of 4861~\AA{}, (for $z = 0$) to a  maximum of 10000~\AA{} (corresponding to $z \simeq 1$). Specifically, 666 sources are used in the CEx diagram, 471 in the MEx diagram, 548 in the [\ion{O}{II}] BPT diagram (`blue' diagram), 104 in the [\ion{N}{II}] BPT diagram and 106 in the [\ion{S}{II}] BPT diagram.  Also, 100 of these 726 sources have matched x-ray counterparts from the list of \cite{2017_Luo}. All this information is summarized in~\autoref{data_reduction} which provides an overview of the data selection and analysis workflow.

A comment on reddening effects is also important here. In our study, galactic reddening is negligible ($N_\mathrm{H} \sim 8 \times 10^{19}\ \mathrm{cm}^{-2}$), as the GOODS-South field lies in a region of low galactic hydrogen column density and dust extinction \citep{1998_Schlegel,2004_Giavalisco}, making the effect of foreground reddening minimal. To assess the presence of intrinsic reddening (in the host), we analyze the Balmer decrement \citep{2010_Grupe,2008_Dong}, \textit{i.e.} the H$\alpha$/H$\beta$ ratio. Following the criterion proposed by \citet{2008_Dong} and also adopted by \citet{2010_Grupe}, sources with H$\alpha$/H$\beta < 3.06$ are considered to exhibit no significant intrinsic reddening. Out of the 726 sources used in our diagnostic diagrams, 125 lie at redshifts where both H$\alpha$ and H$\beta$ fall within the observed spectral range, allowing for flux measurements for both lines. Among these, 37 sources ($\sim$30\%) do not show significant broad-line H$\alpha$ or H$\beta$ emission, likely due to weak BLR activity. Of the remaining sources, 82 ($\sim$65\%) show H$\alpha$/H$\beta$ ratios below 3.06, consistent with little to no intrinsic reddening. Only 6 sources ($\sim$5\%) have H$\alpha$/H$\beta > 3.06$, suggesting the presence of intrinsic reddening in their BLR and the need for reddening correction. This analysis indicates that, statistically, a vast majority ($\sim$95\%) of the sources in the data considered by us would not exhibit significant intrinsic reddening. Although the number of sources with available Balmer line measurements is limited, the sample of 125 is statistically sufficient to infer overall trends. Thus, we opine that reddening effects would not play a significant role in the broader interpretation of statistical trends in the present GOODS-south dataset.


\begin{table}
    \centering
    \caption{List of redshift surveys used}
    \label{redshift_surveys}
    \begin{tabular}{lc} 
    \hline
    \textbf{Redshift Survey} & \textbf{Number of Sources} \\
    \hline
    \cite{2005_Vanzella_AandA} [V05] & 234 \\
    \cite{2006_Vanzella_AAP} [V06] & 723 \\
    \cite{2007_Ravikumar_AAP} [R07] & 156 \\
    \cite{2008_Vanzella_AAP} [V08] & 886 \\
    \cite{2010_Balestra_AAP} [B10] & 2824 \\
    \cite{2014_Hsu_APJ} [H14] & 261 \\
    \cite{2015_Morris_AJ} [M15] & 305 \\
    \cite{2021_Joshi_APJ} [J19] & 154 \\
    \hline
    \end{tabular}
\end{table}

\begin{table}
    \centering
    \caption{Data Reduction Stages and Source Statistics}
    \begin{tabular}{@{}llr@{}}
        \hline
        \textbf{Stage} & \textbf{Description} & \textbf{Count} \\ 
        \hline
        Stage 1 & Total ESO Spectra & 8864 \\ 
        \hline
        Stage 2 & Total Redshifts Available & 4845 \\ 
        \hline
        Stage 3 & Total Sources with Unique Redshifts & 3534 \\ 
        \hline
        Stage 4 & Total Fitted Spectra & 2937 \\
        & (Includes sources with only Host Galaxy Fitting also)  \\ 
        \hline
        Stage 5 & Total Unique Sources used in diagnostic diagrams & 726\\ 
        & (Considering sources with at least (\ion{O}{III}/HB) ratio)  \\ 
                & Sources in CEx Diagram & 666 \\ 
                & Sources in MEx Diagram & 471 \\ 
                & Sources in [\ion{O}{II}] BPT Diagram & 548 \\ 
                & Sources in [\ion{N}{II}] BPT Diagram & 104 \\ 
                & Sources in [\ion{S}{II}] BPT Diagram & 106 \\ 
                & Matched X-ray Sources (from \citep{2017_Luo}) & 100 \\
        \hline
    \end{tabular}
    \label{data_reduction}
\end{table}

\section{Spectral Fitting Methodology}

\label{Fitting Section}

\begin{figure*}
    \centering
    \includegraphics[width=1\linewidth]{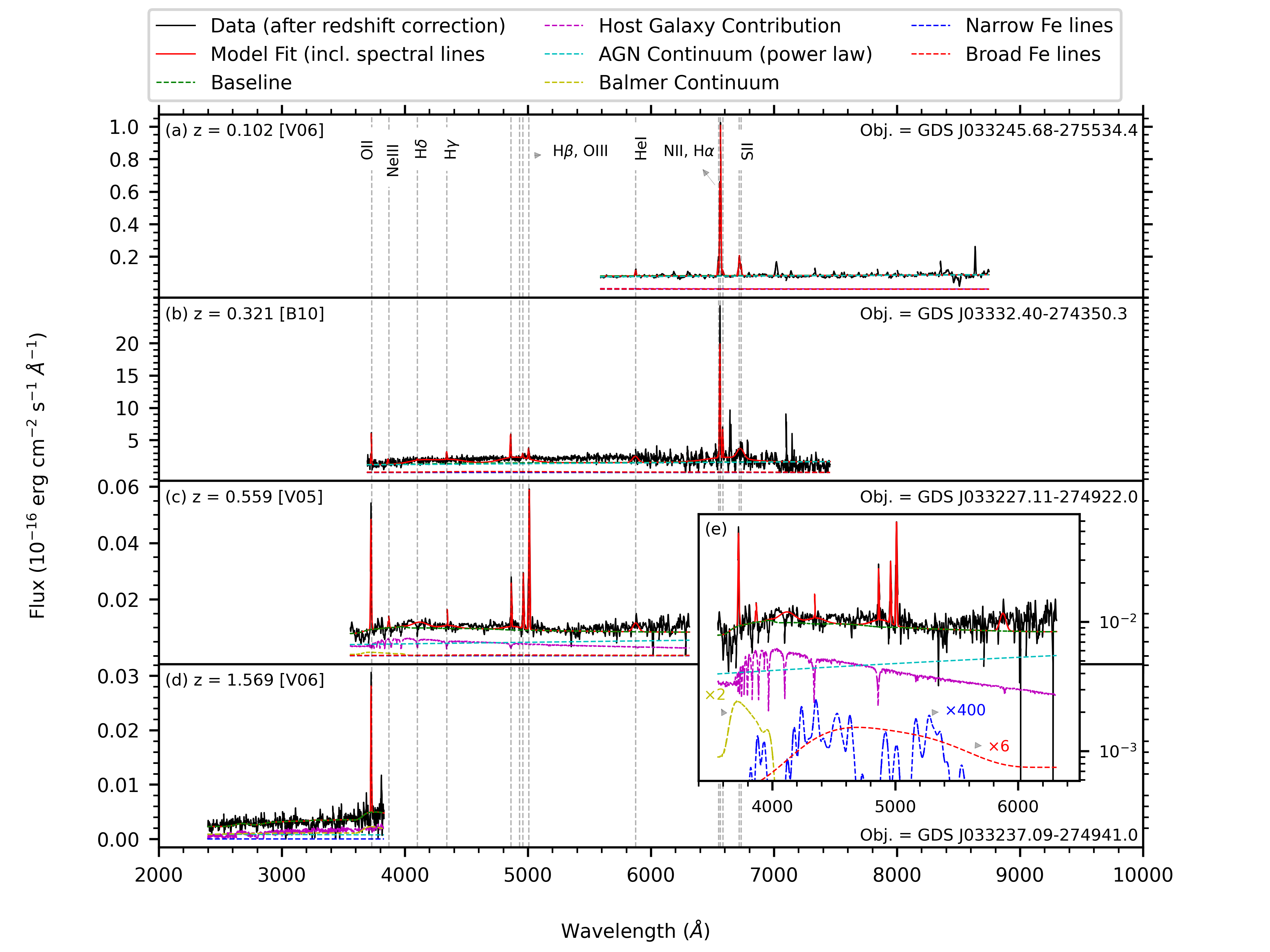}
    \caption{Spectral model fitting results for four sources: (a) \textit{GDS J033245.68-275534.4} ($z=0.102$), (b) \textit{GDS J033332.40-274350.3} ($z=0.321$), (c) \textit{GDS J033227.11-274922.0} ($z=0.559$), (d) highlights \textit{GDS J033237.09-274941.0} ($z=1.569$). Across the subfigures, we have shown the redshift-corrected spectra (black curve) and the overall model fit (red curve, including discrete emission lines), as well as baseline contributions, \textit{i.e.} host galaxy contribution (magenta), AGN power-law continuum (cyan), and narrow and broad Fe lines (blue and red dashed). Subfigure (e): log-scaled version of (c), emphasizing all the spectral components for clarity.}
    \label{fig_1}
\end{figure*}

There are a few spectral fitting codes for galaxy / AGN spectra already mentioned in the literature~\citep{2011_Shen_ApJS, 2017_Calderone_MNRAS, 2023_Ilic_ApJ_Fantasy}. Early tools like the SPECFIT package \citep{1994_Kriss} provided foundational capabilities for continuum fitting and basic emission line modeling but lacked comprehensive integration of multi-component spectral features. A major leap was made by \citet{2011_Shen_ApJS}, who performed a systematic analysis of over 100,000 SDSS quasar spectra, including detailed measurements of major emission lines and continua. Subsequent efforts, such as QSFit by \citet{2017_Calderone_MNRAS}, introduced frameworks for global fitting of continuum, absorption, and emission features. More recently, the FANTASY code \citep{2023_Ilic_ApJ_Fantasy} demonstrated high-fidelity multicomponent fitting for AGN spectra, employing broken power-law continua and refined \ion{Fe}{II} models derived from eigenspectra. However, there are also constraints associated with them, \textit{e.g.} some spectral features are not included, or they are usable only for datasets from specific observatories, etc. Building upon these insights, we have indigenously developed a Python-based spectral fitting code \citep{2024asi..confP..53P}, and utilised it for the spectral analysis of GOODS-S optical spectra. The detailed algorithm is described in the Appendix. Here, we focus on more physical aspects of the code. 

Firstly, the continuum emission from the galaxy is considered to be comprised primarily of four distinct components of different physical origins. The continuum emission from the (active) galactic nucleus is modelled as a broken power law (e.g. see~\cite{2017_Calderone_MNRAS}). A novel feature of our code is the inclusion of a host galaxy contribution model, achieved through templates from the CB2019 library developed by Bruzual and Charlot~\citep{2003_Bruzual_Charlot_MNRAS, 2020_Coelho_MNRAS_CB2019}, which provides spectral evolution data for stellar populations ranging from 0.1 Myr to 14 Gyr across a metallicity range of 0.0001 to 0.06. The host galaxy fitting process involves including the galaxy templates as a part of the baseline, and applying a least-squares fitting algorithm by iterating across all templates in the library to identify the best-fitting template. In the baseline continuum, we also consider the Balmer jump, or Balmer discontinuity, which occurs near the Balmer limit (3646 \AA), with the Balmer continuum modelled using a blackbody function~\citep{1985_Wills_APJ, 1995_Storey_MNRAS, 2017_Calderone_MNRAS}. The model represents the continuum effectively below the Balmer edge wavelength, while suitably broadened Gaussian functions are used for the Balmer lines above it. Additionally, the model incorporates a pseudo-continuum due to both broad and narrow Fe lines, which must be included in estimating the overall continuum and addressing contamination (\textit{e.g.} from lines like H$\beta$). This pseudo-continuum is generated by applying suitable Gaussian broadening to the iron lines in the UV \citep{2001_Vestergaard_Wilkes_APJS} and optical \citep{2004_Veron-Cetty_AandA} wavelength ranges. Thermal broadening is a critical parameter in this process, with specific bounds established for narrow and broad lines in both regions, ensuring the physical validity and reliability of the fitting.

The discrete emission lines rising above the aforementioned continuum (baseline) typically include \ion{O}{II}, \ion{Ne}{III}, H$\delta$, H$\gamma$, H$\beta$, H$\alpha$, \ion{O}{III}, \ion{He}{I}, \ion{N}{II}, and \ion{S}{II} lines. The fitting of these lines is accomplished using Gaussian functions that model these lines based on parameters such as amplitude, FWHM, and velocity offset. Specifically, FWHM values for narrow lines range from 100-2000 km s$^{-1}$ and for broad lines from 1000-15000 km s$^{-1}$, while velocity offsets are constrained to $\pm 1000$ km s$^{-1}$ for narrow lines and $\pm 3000$ km s$^{-1}$ for broad lines. These bounds have been successfully used in the past for galaxy / AGN spectral analysis (see \textit{e.g.} \cite{2017_Calderone_MNRAS}). Overall, our model attempts to reproduce all the observed spectral features based on physically motivated model components.

\section{Results and Discussion }

\label{Results and Discussion}

\autoref{fig_1} shows the fitting results for four sources across different wavelength ranges (after redshift correction), showcasing many of the emission lines typically considered in this study. \autoref{fig_1} (a) displays the object \textit{GDS J033245.68-275534.4} with a redshift of 0.102 [V06]. The solid black curve represents the data after redshift correction, while the solid red curve indicates the model fit, including spectral lines. Contributions from the host galaxy, AGN continuum (broken power law), and narrow and broad Fe lines are shown as dashed magenta, cyan, blue and red lines respectively, forming an overall baseline (green dashed line), upon which prominent emission lines, such as [\ion{O}{III}]$\lambda$5007, H$\beta$, H$\alpha$, etc. are fitted. Similarly, subfigures (b), (c) and (d) show the objects \textit{GDS J033332.40-274350.3}, \textit{GDS J033227.11-274922.0} and \textit{GDS J033237.09-274941.0}, with progressively increasing redshifts of 0.321 [B10], 0.559 [V05] and 1.569 [V06]. The inset in \autoref{fig_1} (e) provides a log-scaled view of the same spectrum as in \autoref{fig_1} (c), emphasising specific spectral contributions. The same spectral fitting algorithm is used to fit all 4845 sources with known redshift values, achieving a decomposition of different components contributing to the spectral features in these sources~\cite{2024asi..confP..53P}.

\subsection{[\ion{O}{II}], [\ion{N}{II}], and [\ion{S}{II}] BPT diagrams \label{BPTs_trad}}

\begin{figure*}
    \centering
    \includegraphics[width=1\linewidth]{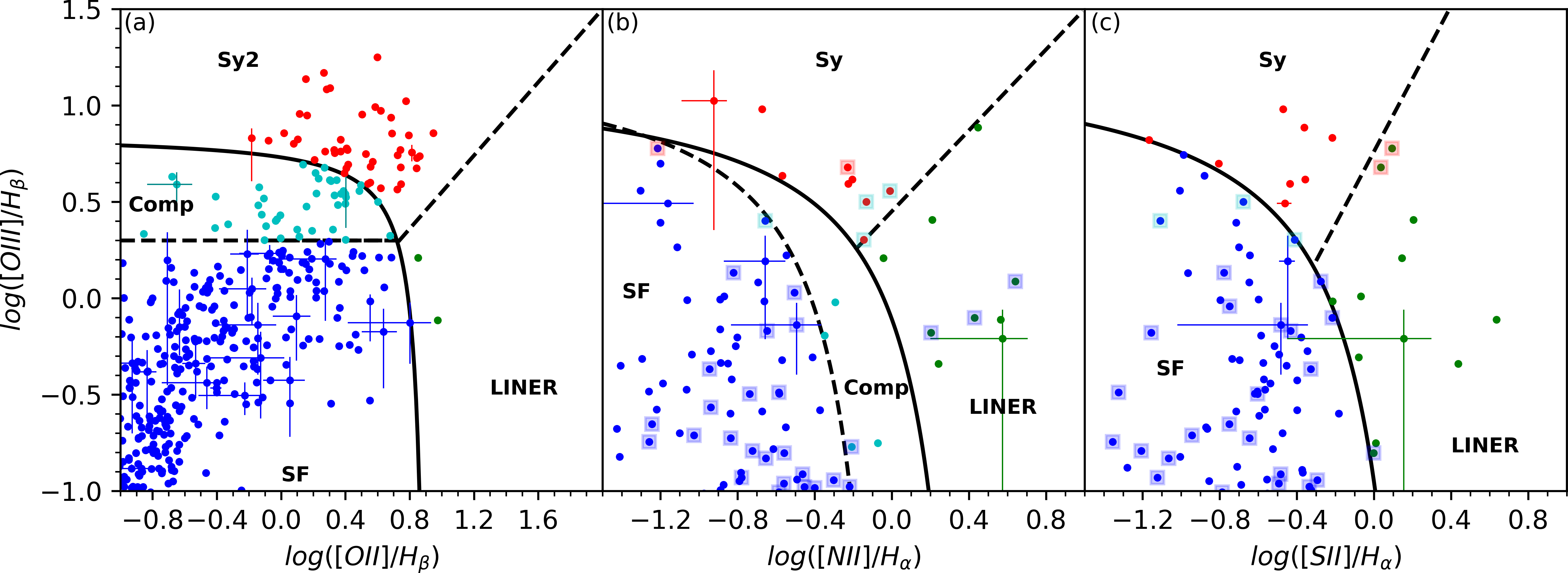}
    \caption{(a) The [\ion{O}{II}] BPT diagram (`blue diagram'), (b) The [\ion{N}{II}] BPT diagram, (c) The [\ion{S}{II}] BPT diagram. Across subfigures, blue dots represent star-forming galaxies (SFs), cyan dots indicate composite galaxies, red dots correspond to Seyferts (AGN), and green dots denote LINERs (AGN). Square markers (blue, cyan, red, and green) indicate classifications from the [\ion{O}{II}] BPT diagram, providing a comparison across subfigures.}
    \label{fig_2}
\end{figure*}

The classification of extragalactic sources into AGNs and SFs can be effectively achieved using various BPT diagrams, which are plotted together in \autoref{fig_2}, for the GOODS-S data considered in this paper.

\autoref{fig_2} (a) presents the [\ion{O}{II}] BPT diagram (the so-called `blue diagram'~\citep{2010_Lamareille}), which compares the ratios [\ion{O}{III}]$\lambda$5007/H$\beta$ and [\ion{O}{II}] $\lambda$3727/H$\beta$. Most sources are located in the bottom-left region, identified as SF galaxies (blue dots). The upper and right regions are dominated by AGNs, including Seyfert galaxies (red dots) and LINERs (green dots). There is a region where the star-forming and Seyfert populations overlap, and the classification for these sources is uncertain. This is the so-called composite region (cyan dots). The empirical lines shown are adapted from \cite{2010_Lamareille}. 

Along the same lines, \autoref{fig_2} (b) shows the [\ion{N}{II}] BPT diagram which plots [\ion{O}{III}]$\lambda$5007/H$\beta$ against [\ion{N}{II}]$\lambda$6584/H$\alpha$. This diagnostic separates SF galaxies (blue dots) from AGNs (red and green dots for Seyferts and LINERs, respectively) and includes a composite region (cyan dots) for sources with ambiguous classifications. Square markers (blue, cyan, red, and green) indicate classifications from the [\ion{O}{II}] diagram, revealing some discrepancies between the two methods, such as SF sources classified as AGNs and vice versa. The dividing lines are based on \cite{2003_Kauffmann_MNRAS_a} and \cite{2007_Schawinski_MNRAS}. 

Finally, \autoref{fig_2} (c) displays the [\ion{S}{II}] BPT diagram, which compares [\ion{O}{III}]$\lambda$5007/H$\beta$ and [\ion{S}{II}] $\lambda$$\lambda$6717,6731/H$\alpha$. It effectively distinguishes SF galaxies (blue dots) from AGNs (Seyferts and LINERs in red and green dots, respectively). Similar to \autoref{fig_2} (b), square markers indicate the [\ion{O}{II}] classification. The dividing lines are derived from \cite{2006_Kewley}. Across all subfigures, the majority of sources are concentrated in the SF region, consistent with the general consensus on the global AGN population, as well as previous reports, \textit{e.g.} by \cite{2006_Kewley} and \cite{2011_Juneau_APJ}.

Our dataset also contains some sources for which multiple spectra are available, acquired at different epochs (see~\autoref{Data_section}). For these selected sources, we have the opportunity to quantify the long-term time-variability in the emission lines, which may reflect changes in physical conditions, ionisation states, or other temporal factors in the sources \citep{1988_Peterson, 1993_Peterson}. To represent this variability on the BPT diagram, we plot the observed emission-line flux ratios for each source, with the marker position indicating the mean flux ratio across multiple observations, and the error bars denoting the range of flux ratios observed at different times. These error bars are intended as a measure of the temporal variability, highlighting the dynamic nature of the emission-line regions, and the impact of this variability on the corresponding classification scheme. While AGNs exhibit variability due to fluctuations in their accretion rates and surrounding environments, and SF galaxies may undergo changes in their star formation activity, it is observed that the BPT classification remains largely unaffected. This is because the emission lines used in these diagrams originate from ionized gas regions that respond to long-term radiation fields rather than short-term fluctuations. Thus, irrespective of variability in AGNs or star-forming galaxies, BPT diagrams remain effective and reliable for distinguishing between different galaxy types. 

Across the different BPT diagnostic diagrams, the classification statistics reveal consistent and well-separated source populations. In the [\ion{O}{II}] BPT, a total of 459 sources are classified as SF ($\sim$84\%), 41 as composite ($\sim$7\%), and 48 as AGN ($\sim$9\%). Similarly, the [\ion{S}{II}] diagram identifies 84 SF and 22 AGN sources ($\sim$79\% and $\sim$21\%, respectively); while the [\ion{N}{II}] diagram yields 77 SF, 8 composite, and 19 AGN sources ($\sim$74\%, $\sim$8\% and $\sim$18\%, respectively). 

When comparing [\ion{O}{II}] classifications with those from [\ion{S}{II}] and [\ion{N}{II}], there is a strong level of agreement. Among the 459 SF sources in [\ion{O}{II}], 28 are also present in the [\ion{S}{II}] diagram, of which 27 ($\sim$96\%) are classified as [\ion{S}{II}]-SF. The corresponding numbers for [\ion{N}{II}]-SF are 28 out of 33 ($\sim$85\%). In the case of [\ion{O}{II}]-identified composite and AGN sources, only 2 sources each are present in the [\ion{S}{II}] and [\ion{N}{II}] diagrams, and hence are statistically insufficient to define any concrete trends. Overall, the [\ion{O}{II}] BPT is found to be quite consistent with [\ion{S}{II}] and [\ion{N}{II}], particularly for the SF population, and is able to cover a much larger number of sources, due to its applicability up to larger redshifts.

\begin{figure}
    \centering
    \includegraphics[width=1\linewidth]{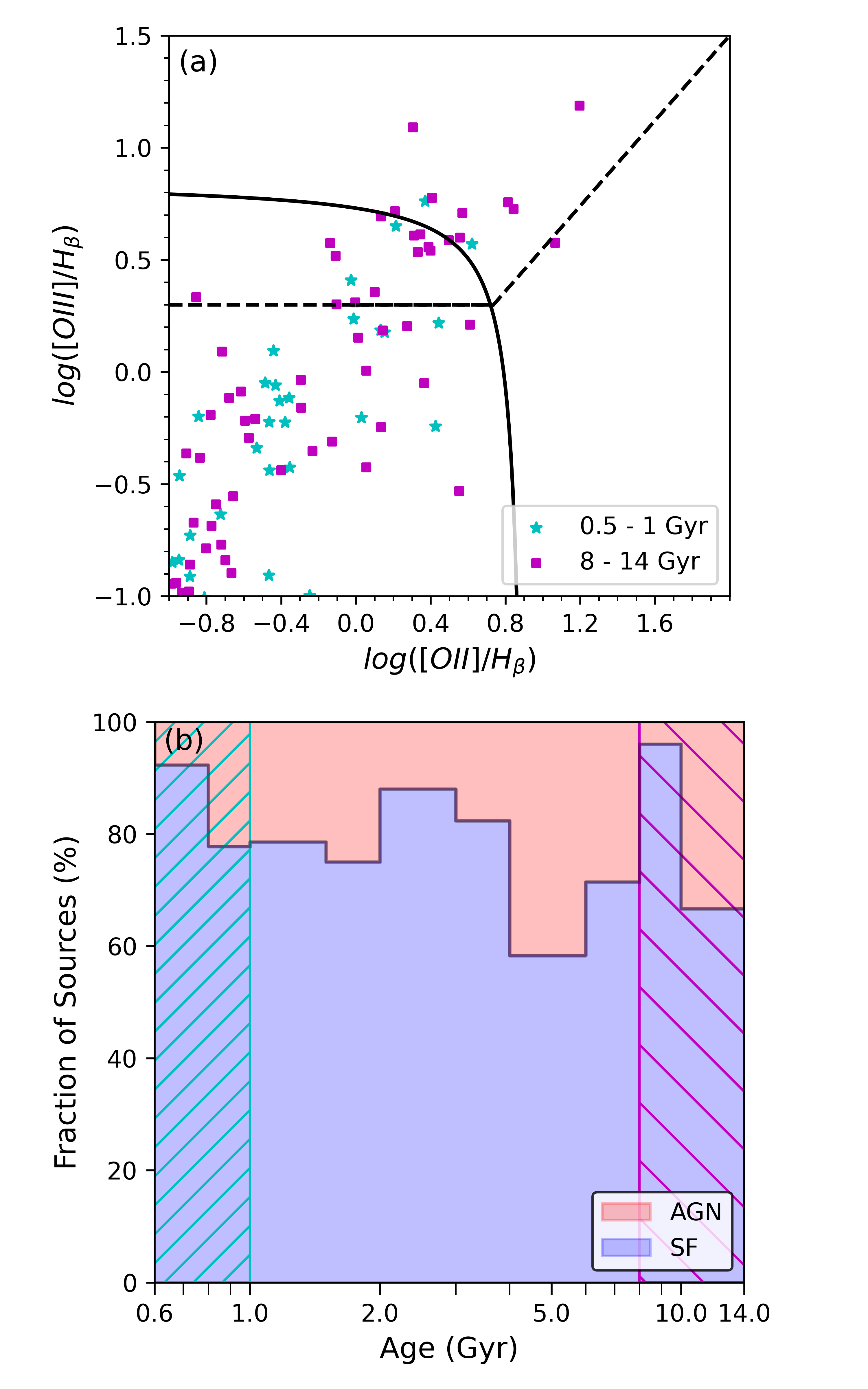}
    \caption{(a): [\ion{O}{II}]-BPT diagram of \autoref{fig_2}(a), limited to sources in two fitted age ranges — 0.5--1~Gyr (cyan stars) and $>$8~Gyr (magenta squares) — representing relatively young and old stellar populations, respectively. While AGNs are predominantly found in the older age bin, a small number of AGNs also appear in the younger age range \citep{2014_Heckman, 2003_Kauffmann_MNRAS_a}. (b): Fractional distribution of [\ion{O}{II}]-classified star-forming galaxies (blue) and AGNs (red) as a function of fitted host galaxy age. A clear trend is seen where the fraction of AGNs increases with stellar age. The two age bins shown in (a) are indicated by shaded boxes for reference.}
    \label{fig_3}
\end{figure}

As discussed in~\autoref{Fitting Section}, the host galaxy contribution is integrally included in our spectral fitting procedure. This uses the stellar population synthesis models of \citet{2003_Bruzual_Charlot_MNRAS}, which predict the spectral evolution of stellar populations over a wide range of ages ($10^5$ to $2 \times 10^{10}$ years) and metallicities. Their high-resolution model incorporates updated stellar evolution tracks and a modern empirical spectral library, enabling the generation of synthetic spectra at \textit{3~\AA} resolution in the optical range (3200--9500~\AA). These models are widely used in galaxy evolution studies for interpreting spectra and deriving star formation histories, metallicities, and dust content \citep{2003_Kauffmann_MNRAS_a,2005_Gallazzi,2011_Popescu}.

We use this library in our code to incorporate the host galaxy component, with the best-fit template providing the galaxy’s age. \autoref{fig_3} presents some results from this analysis. \autoref{fig_3} (a) shows the same [\ion{O}{II}]-BPT diagram as~\autoref{fig_2}, but restricted to sources in two age bins: 0.5--1~Gyr (cyan stars) and $>$8~Gyr (magenta squares), representing relatively young and old systems, respectively. Despite the limited number of sources, a clear pattern emerges: the older population of galaxies are well-represented across different regions of the diagram, while younger galaxies are virtually confined to the [\ion{O}{II}]-SF and [\ion{O}{II}]-composite regions. This is as expected for these age-wise populations, and also yields confidence in our incorporation of galaxy age in our model.

Further, \autoref{fig_3} (b) summarizes the age distribution of [\ion{O}{II}]-classified SFs and AGNs across discrete host galaxy age bins. Each bin corresponds to the intervals used in \autoref{fig_3} (a), marked for consistency. For each age bin, the relative fraction of SFs and AGNs is shown, allowing a direct comparison. A clear trend is observed: the AGN fraction increases with host galaxy age, suggesting a possible evolutionary link between star formation activity and AGN triggering over time.

In each host galaxy age bin, The two highlighted age regions from \autoref{fig_3} (a) are marked in \autoref{fig_3} (b) for reference. It is clearly seen that the fraction of AGNs out of all sources at a given age increases with host galaxy age. In fact, over the complete dataset, if we consider all [\ion{O}{II}]-SF sources, they are uniformly distributed across the age bins ($\sim$6.5\%), whereas the population of all [\ion{O}{II}]-AGNs is significantly skewed towards higher HG ages, rising from $\sim$3\% at $\leq$1 Gyr to $\sim$12\% at $\geq$8 Gyr (not shown in figure).

This consistency between spectral classification and host galaxy age strengthens confidence in our fitted age values. The observed trend underlines an evolutionary connection between star formation and AGN activity: as galaxies age, AGN activity becomes more dominant, potentially due to gas depletion, black hole growth, and central fueling mechanisms \citep{2008_Hopkins,2014_Heckman,2020_Suh}. Meanwhile, the wide age spread of SF galaxies indicates sustained star formation across a large fraction of cosmic time, likely regulated by gas accretion and feedback processes \citep{2014_Madau}. The increased fraction of AGN galaxies at older ages is also understandable in terms of AGN feedback quenching star formation in evolved systems \citep{2012_Fabian,2013_Kormendy}.

\subsection{MEx and CEx diagrams \label{MEx_CEx}}

In addition to the original BPT diagnostic diagrams discussed in~\autoref{fig_2}, there have been recent efforts to use the same framework, modified by the inclusion of some other physical parameters. Examples of such efforts in the literature are the MEx~\citep{2011_Juneau_APJ} and CEx~\citep{2011_Yan_APJ} diagrams, which utilise the stellar mass $M^*$ and the $(U-B)$ colour index respectively.

The MEx diagram, originally developed by \cite{2011_Juneau_APJ}, plots the stellar mass of the galaxies (\(\log(M_*/M_\odot)\)), against the \([\text{\ion{O}{III}}]/\text{H}\beta\) flux ratio, which reflects the ionization state of the gas. The diagram is divided into regions corresponding to different galaxy types: SF, composite, and AGN-dominated galaxies, separated by solid and dashed empirical lines adapted from \cite{2014_Juneau}. The original demarcations in \citep{2011_Juneau_APJ} were updated by \cite{2014_Juneau} to account for emission line detection limits, broadening the diagram's applicability across redshifts. 

To obtain the stellar masses ($M^*$) we have used the work of \cite{2015_Santini_APJ}, which summarises the efforts by 10 different teams within the CANDELS collaboration, who compute stellar masses using the CANDELS-GOODS-S dataset. For our analysis, we adopt stellar masses computed by the mass estimation method 2d$_\tau$, which assumes a Salpeter Initial Mass Function (IMF). This choice is mainly to ensure consistency with the AGN classification criteria in the MEx diagram developed by \cite{2011_Juneau_APJ,2014_Juneau}.

The physical motivation and relevance of the MEx diagram has been emphasised in the literature~\citep{2011_Juneau_APJ, 2013_Vitale}, and is summarised below. Galactic stellar mass is an important parameter as it correlates with the overall gravitational potential and metallicity of the galaxy, influencing the ionisation state of the gas. Similarly, the [\ion{O}{III}]$\lambda$5007/H$\beta$ ratio is sensitive to both the ionisation state and the metallicity. AGNs typically exhibit higher [\ion{O}{III}]$\lambda$5007/H$\beta$ ratios due to the harder ionising radiation from the AGN compared to the softer radiation from young stars in star-forming galaxies. Galaxies with higher stellar masses are more likely to host AGNs, as these are often found in more massive galaxies with older stellar populations and higher central concentrations, which favor AGN activity~\citep{2013_Vitale}. 

\begin{figure}
    \centering
    \includegraphics[width=1\linewidth]{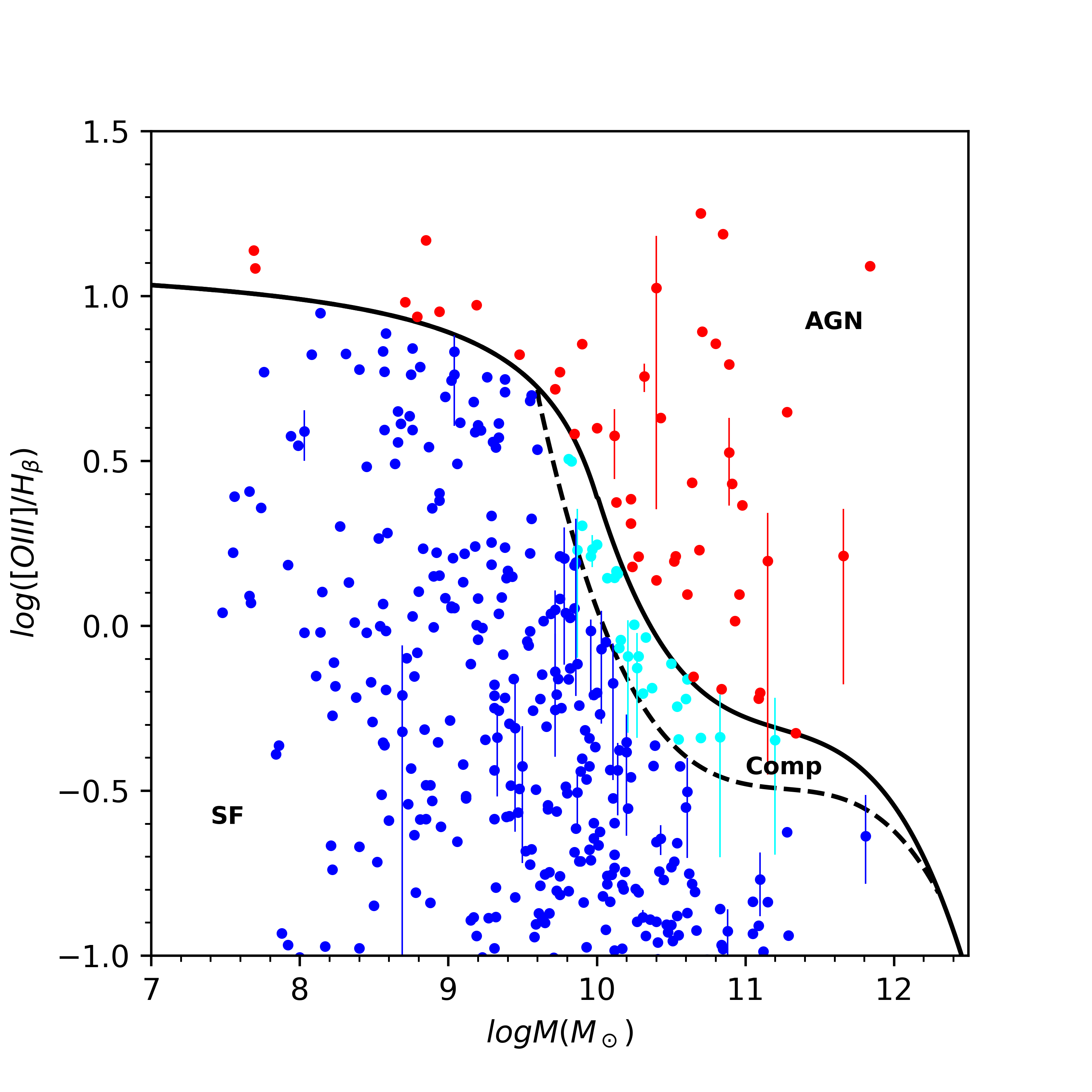}
    \caption{MEx diagram for the present data. The bottom-left region indicates star-forming galaxies (SF, blue dots), the composite region is located between the solid and dashed lines (Comp., cyan dots), and the top-right region corresponds to AGN activity (AGN, red dots). The empirical lines are adapted from \protect\cite{2014_Juneau}.}
    \label{fig_4}
\end{figure}

In the MEx diagram for the data considered by us (see~\autoref{fig_4}), 400 sources are classified as SF ($\sim$85\%). These are located in the bottom-left region and are represented by blue dots. The middle (composite) region contains 26 galaxies ($\sim$5\%), whose classification is ambiguous, marked by cyan dots. The top-right region comprises 45 sources classified as AGNs ($\sim$10\%), indicated with red dots. Towards the end of this section, we present a comparison of the MEx results with our earlier results from~\autoref{BPTs_trad}.



Apart from the MEx diagram, the CEx diagram is another diagnostic tool used to classify galaxies based on their spectral and photometric properties, particularly for distinguishing between SF galaxies, composite systems, and AGN-dominated sources. Developed by~\cite{2011_Yan_APJ}, the diagram employs the [\ion{O}{III}]$\lambda$5007/H$\beta$ flux ratio and the rest-frame $(U-B)$ color index to distinguish between these classes, with empirical boundaries (black solid and dotted lines) separating different regions. Star-forming galaxies occupy the region with low [\ion{O}{III}]$\lambda$5007/H$\beta$ values and bluer $(U-B)$ colors, due to ionization dominated by young, massive stars. Composite galaxies lie in an intermediate zone with moderate values of both parameters, indicating contributions from both star formation and AGN activity. AGN-host galaxies, with high [\ion{O}{III}]$\lambda$5007/H$\beta$ and redder $(U-B)$, are located in the upper-right region, where ionization is primarily driven by accretion onto supermassive black holes. 

\cite{2013_Trump} later refined the AGN/SF division (grey dotted line) by recalibrating the empirical boundaries using a larger and more diverse galaxy sample, including sources at $z > 1$. This makes the CEx diagram particularly effective in cases where traditional BPT diagrams are limited, such as for galaxies lacking all necessary emission lines.

Physically, the diagram is founded on the observation that nearly all BPT-identified AGNs reside in red or green galaxies, while very few are found in blue ones~\citep{2011_Yan_APJ}. Blue galaxies typically have lower stellar masses and bulge-to-disk ratios, which correlate with smaller central black holes (BHs)~\citep{2004_Haring,2011_Jahnke}. Consequently, AGNs in such galaxies tend to be less luminous and more easily overwhelmed by star formation, making them harder to detect using standard BPT diagnostics~\citep{2013_Trump,2014_Juneau}. The CEx method attempts to correct for this bias by incorporating galaxy color, allowing for more reliable AGN identification across the full color sequence.

In our analysis, we use $(U-B)_0$ values from the COMBO-17 survey~\citep{2004_Wolf_AAP}, as described in~\autoref{Data_section}. Since the COMBO-17 magnitudes are in the Vega system and the CEx diagnostic was calibrated using AB magnitudes~\citep{2011_Yan_APJ}, we converted the $(U-B)$ colors to the AB system using the correction provided by~\cite{2006_Willmer}.

\begin{figure}
    \centering
    \includegraphics[width=1\linewidth]{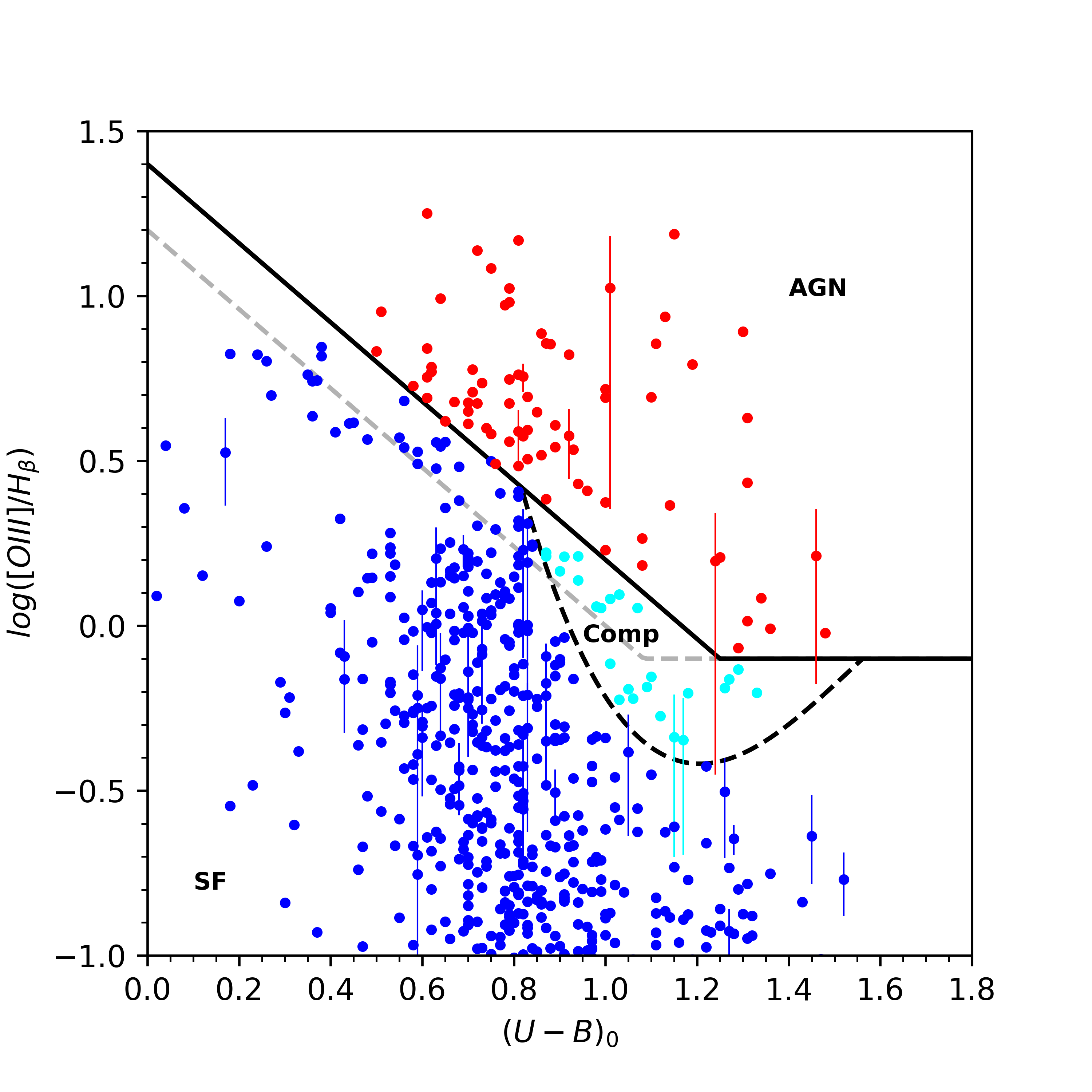}
    \caption{CEx diagram for the present data. Star-forming galaxies (SF, blue dots) occupy the region with low [\ion{O}{III}]$\lambda$5007/H$\beta$ ratios and bluer \(U-B\) colors, composite galaxies (Comp., cyan dots) are located in the intermediate region, and AGN-host galaxies (AGN, red dots) are found in the region with high [\ion{O}{III}]$\lambda$5007/H$\beta$ ratios and redder \(U-B\) colors. The empirical boundaries are adapted from \protect\cite{2011_Yan_APJ}.}
    \label{fig_5}
\end{figure}

In the CEx diagram (see~\autoref{fig_5}) for the data considered by us, 569 sources are classified as star-forming (SF; $\sim$85\%). These galaxies, represented by blue dots, predominantly belong to the blue, low-mass population, where ionization is primarily attributed to young, massive stars. The diagram also identifies 23 composite galaxies ($\sim$4\%), marked by cyan dots, whose classification is ambiguous (\textit{e.g.} see \cite{2011_Yan_APJ}). Finally, 74 sources ($\sim$11\%) are classified as AGNs and are shown in red. These galaxies typically lie on the redder end of the color sequence and host more massive black holes, where ionization is expected to be dominated by AGN accretion activity. The CEx diagram thus offers a valuable classification scheme, especially for galaxies where the BPT diagnostics are limited or unavailable.

\begin{figure*}
    \centering
    \includegraphics[width=\textwidth]{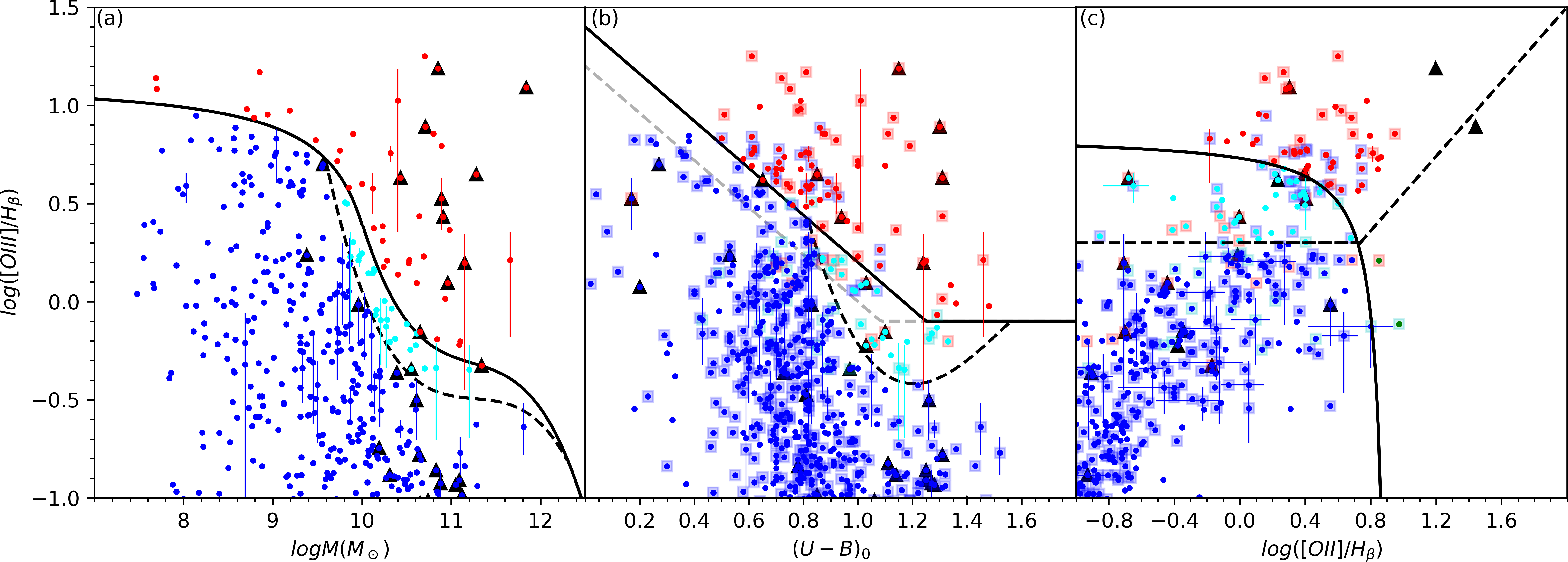 }
    \caption{The figure presents source classifications across the MEx, CEx, and [\ion{O}{II}] BPT diagnostic diagrams. Subfigures are color-coded to indicate star-forming galaxies (blue), composites (cyan), and AGNs (red). (a) shows the MEx diagram, where AGN identification is compared to X-ray-selected sources. (b) displays the CEx diagram, with classifications also represented as shaded squares for comparison. (c) presents the [\ion{O}{II}] BPT diagram, illustrating classification variations and the challenges of distinguishing AGNs from star-forming and composite galaxies. Sources identified as X-ray AGNs from the \protect\cite{2017_Luo} CDF-S catalog are marked as black triangles, providing insight into how different diagnostics classify these objects.}
    \label{fig_6}
\end{figure*}

In \autoref{fig_6}, we examine the behavior of sources across various diagnostic diagrams, \textit{viz.}, in the sequence shown in the figure, the MEx (a), CEx (b), and [\ion{O}{II}] BPT (c) diagnostic diagrams. The subfigures are colour-coded such that classification of galaxies into SF, composite, and AGN-dominated categories is indicated by the colour of the markers (dots). In subfigures (b) and (c), the MEx classification is additionally shown in the form of shaded squares. Furthermore, sources identified as X-ray AGNs by \cite{2017_Luo} from the CDF-S source catalog are also shown as black triangles.

In \autoref{fig_6} (a), the MEx diagram shown is the same as in~\autoref{fig_4}, and classifies sources into SFs (blue dots), composites (cyan dots), and AGNs (red dots). Additionally, X-ray AGNs are marked as black triangles, showing that 11 out of 32 X-ray AGNs are correctly identified as MEx-AGNs, while 20 are classified as MEx-SFs, and 1 as a MEx-Composite. 

\autoref{fig_6} (b) shows the CEx diagram (same as~\autoref{fig_5}), with the same colour-coding of markers (dots). In addition, the MEx classification is also overlaid in the form of blue squares (MEx-SFs), cyan squares (MEx-Composites), and red squares (MEx-AGNs), for comparison. In the CEx classification, out of 35 X-ray AGNs, 7 are identified as CEx-AGNs, 25 as CEx-SFs, and 3 as CEx-Composites (accuracy $\sim$20\%).

\autoref{fig_6} (c) displays the [\ion{O}{II}] BPT diagram (see~\autoref{fig_2}(a)), where blue, cyan, and red squares represent \ion{O}{II}-SFs, \ion{O}{II}-Composites, and \ion{O}{II}-AGNs, respectively, and the MEx classification is overlaid for comparison similar to \autoref{fig_6} (b). Out of 30 X-ray AGNs, only 3 are classified as \ion{O}{II}-AGNs, with the majority falling into the \ion{O}{II}-SF and \ion{O}{II}-Composite categories. 

In broader terms, misclassifications of X-ray AGNs into SF or composite categories highlight the limitations of all these diagnostic methods. To summarise our results, the MEx diagram provides better AGN identification among X-ray-identified AGNs compared to other diagrams, correctly classifying $\sim$35\% (compared to $\sim$20\% and $\sim$10\% for the CEx and [\ion{O}{II}]-BPT, respectively). This observation is consistent with the analysis by \cite{2017a_Ricci, 2017b_Ricci, 2018_Hickox}, in which they conclude that $\sim$70\% of AGNs exhibit obscuration to varying degrees, making them easier detected in the X-ray spectrum than in the optical. These results emphasize the need for complementary multi-wavelength data to improve AGN detection. 

Using the MEx diagram as a reference, we also assess the consistency of source classifications from the [\ion{O}{II}] and CEx diagrams. Among the 276 sources classified as star-forming in MEx, 264 ($\sim$88\%) are also identified as SF in the [\ion{O}{II}] diagram, and 361 ($\sim$93\%) are consistent with the SF classification in the CEx diagram. In contrast, for the 45 sources classified as AGN/composite in MEx, only 25 ($\sim$68\%) match the AGN/composite category in [\ion{O}{II}], whereas 39 ($\sim$89\%) show consistent AGN/composite classification in CEx. 

These results suggest that the CEx diagram shows stronger correlation with the MEx classification, especially for both SF and AGN/composite sources, while the [\ion{O}{II}] diagram exhibits slightly lower consistency, particularly in the AGN/composite regime. 


\begin{figure*}
    \centering
    \includegraphics[width=\textwidth]{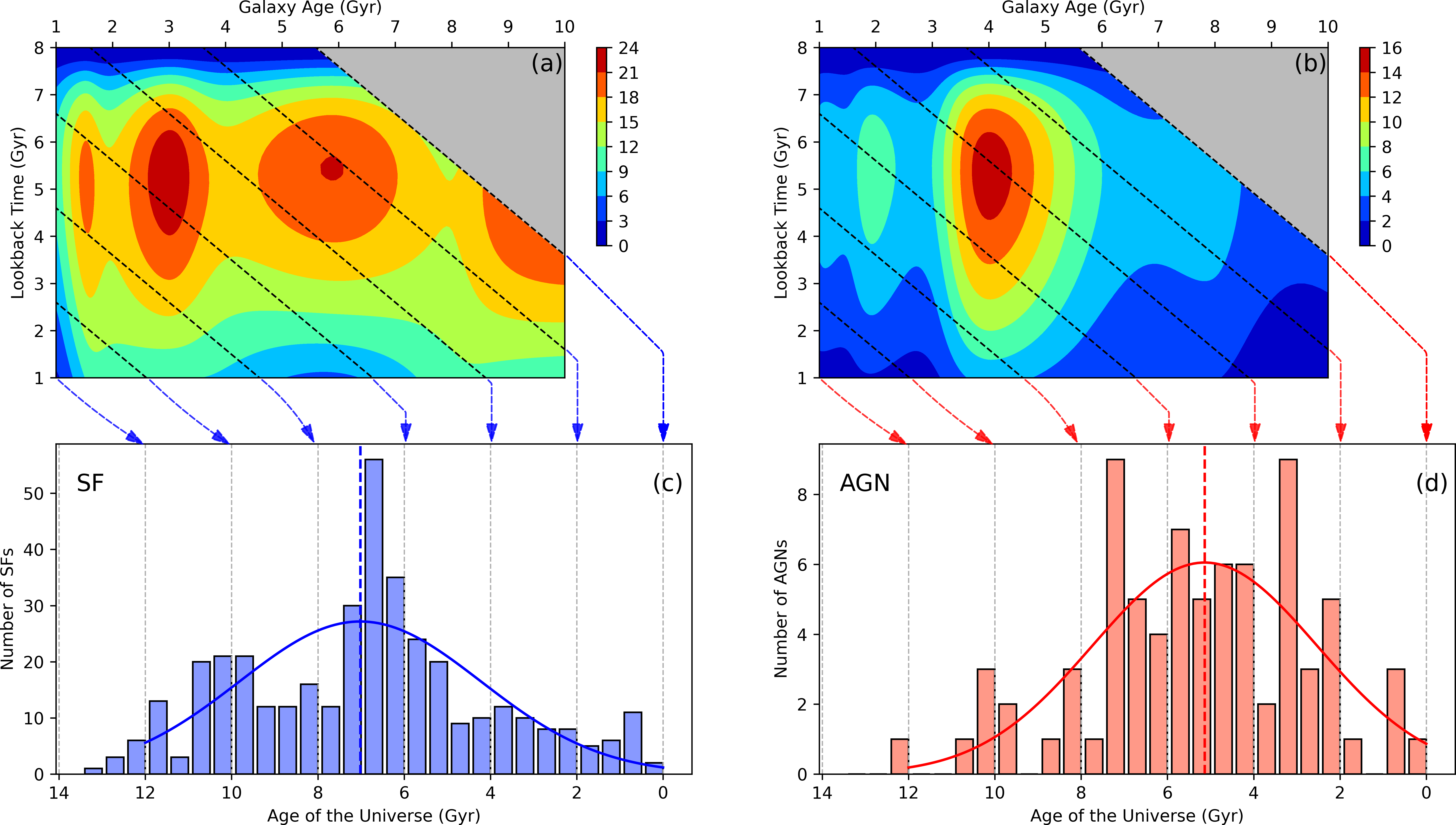}
    \caption{The Lookback Time vs. Host Galaxy Age plot reveals distinct evolutionary trends for star-forming galaxies and AGNs. The peak of AGN activity occurs at earlier cosmic times, suggesting that black hole growth was more significant when the universe was hotter, likely due to enhanced gas accretion. In contrast, star-forming galaxies are observed across all epochs, with their peak activity occurring later, indicating sustained star formation over cosmic time. The distribution of sources shows a clear shift, with AGNs being more prevalent in older host galaxies, while star formation remains widespread. This suggests a transition in galaxy evolution, where AGN feedback may have played a role in regulating star formation in massive systems.}
    \label{fig_7}
\end{figure*}

\subsection{Temporal Evolution: Age vs Lookback time \label{Temporal Evolution}}


In this section, we present a novel use of the host galaxy ages determined as part of the spectral fitting. We combine the derived host galaxy ages with the redshift values ($z$) for the respective galaxies. The redshift provides us with the `lookback time' ($\tau_\text{lookback}$), which indicates how far back in time we are observing the galaxy. The host galaxy age ($\tau_\text{age}$), as derived from the spectra, represents the age of the galaxy at the time of light emission. The sum of these two quantities ($\tau_\text{tot} = \tau_\text{lookback} + \tau_\text{age}$) provides the total time elapsed since the epoch when the respective galaxy originated. This, in turn, acts as a measure of the age of the Universe at that epoch (13.6 Gyr - $\tau_\text{tot}$). 

In simple terms, $\tau_\text{lookback}$ tells us `how long ago' the light was emitted, while $\tau_\text{age}$ tells us `how old' was the galaxy at the time of emission. The lookback time ($\tau_\text{lookback}$) is derived from the redshift using cosmological parameters from the Planck 2018 results \citep{2020_Planck}, adopting $\Omega_m = 0.315$, $\Omega_\Lambda = 0.685$, and $H_0 = 67.4$ km s$^{-1}$ Mpc$^{-1}$. Using these values, the current-epoch age of galaxies is computed as the sum of Lookback Time and Galaxy Age ($\tau_\text{tot} = \tau_\text{lookback} + \tau_\text{age}$). The uniqueness of our methodology is that both time parameters ($\tau_\text{age}$ and $\tau_\text{lookback}$) are derived entirely from the optical spectrum of the given galaxy. Furthermore, $\tau_\text{age}$ is derived from the continuum emission spectrum (host galaxy), while $\tau_\text{lookback}$ is derived from the discrete spectral lines (redshift), and are hence, in principle, independent of each other. Treating them as two independent parameters, one can study the distribution of galaxies across this 2-dimensional data space.

In \autoref{fig_7} (a and b), such a contour map is displayed, showing of the number distibution of galaxies over cosmic time. The horizontal axis represents Galaxy Age $\tau_\text{age}$ (in Gyr), as determined from spectral fitting. The vertical axis represents lookback time $\tau_\text{lookback}$ (in Gyr), as determined from the redshift. By our previous arguments, the diagonal dotted lines are lines of equal universe age ($\tau_\text{tot} = \tau_\text{lookback} + \tau_\text{age}$), and are spaced at intervals of 2 Gyr, starting from 13.6 Gyr on the extreme right, and decreasing towards the bottom left. The coloured contours indicate regions of equal number density of galaxies. The left and right upper subfigures (\autoref{fig_7} (a) and (b)) show this distribution for MEx-SFs and MEx-AGNs respectively. 

The lower subfigures (\autoref{fig_7} (c) and (d)) are histograms generated by diagonally summing the data (\textit{i.e.} summing at equal $\tau_\text{tot}$) from the upper contour plots, as indicated by the dotted-line arrows, but with a bin-width of 0.5 Gyr. Since the diagonal strips represent lines of equal age of the universe, the histogram shows the fraction of galaxies originating at each cosmic epoch. The horizontal scale spans the age of the universe (\textit{i.e.} 13.6 Gyr - $\tau_\text{tot}$, shown increasing from right to left). 

The plots in~\autoref{fig_7} also provide a deeper perspective on the evolutionary timeline of galaxies and their connection to cosmic conditions, which we proceed to discuss below. Our results show that the formation of galaxies which ended up as MEx-SFs peaked when the age of the universe was $7.02 \pm 2.80$ Gyr, whereas MEx-AGNs peaked earlier at $5.14 \pm 2.61$ Gyr. This suggests that AGN formation-conducive activity would have been more prominent at earlier cosmic times. This conclusion is also consistent with previous studies indicating that SMBH growth and AGN activity peaked at earlier cosmic times, likely driven by an abundant cold gas supply and higher merger rates in the early universe \citep{2008_Hopkins, 2014_Heckman}. Furthermore, using standard cosmological parameters, the estimated CMB temperature \citep{2011_Noterdaeme} at universe age 5.14 Gyr (i.e. the peak of AGN-candidate formation in this study) was 5.8 K, while at 7.02 Gyr (the peak of SF-candidate formation), it was 4.7 K. This suggests that a higher cosmic temperature increases the probability of a paticular galaxy to evolve into an AGN, possibly by enhancing gas accretion onto central black holes, thereby triggering AGN-driven feedback (see e.g. \cite{2017_Weinberger}). Conversely, a lower background temperature at later epochs appears to be more conducive to star formation~\citep{2014_Madau}.  

This trend reinforces the idea that AGN activity and star formation are tightly linked to the physical conditions of the Universe at each epoch. The early peak in AGN activity reported by us may have contributed to shaping galaxy evolution, regulating star formation through feedback mechanisms (see e.g. \cite{2012_Fabian, 2013_Kormendy}). Meanwhile, the sustained presence of SF (\textit{i.e} non-AGN) galaxies throughout cosmic history highlights the prolonged role of secular evolution, gas accretion, and internal galaxy dynamics in sustaining star formation \citep{2020_Tacconi}. 

\section{Summary}
  \label{Conclusion}
In this study, we have used the traditional Baldwin-Phillips-Terlevich (BPT) diagram framework ([\ion{N}{II}], [\ion{S}{II}], [\ion{O}{II}] BPT diagrams) to classify AGNs and star-forming galaxies using spectral data from the GOODS-South field. We have also used more modern extensions of the BPT framework (MEx, CEx diagrams) in a detailed comparative study. Our analysis utilized an indigenously developed spectral fitting code, which incorporates host galaxy templates based on the stellar population synthesis models of \cite{2003_Bruzual_Charlot_MNRAS}. This inclusion allows us to extract the the underlying host galaxy continuum from the overall AGN background, and estimate host galaxy ages. This in turn enables a more refined study of AGN and star-forming galaxy evolution.  

Our key findings are as follows:  

\begin{enumerate}  
    \item Traditional BPT diagrams ([\ion{N}{II}], [\ion{S}{II}], [\ion{O}{II}]) remain a relevant classification tool, even in the presence of spectral variability, with the [\ion{O}{II}] BPT diagram serving as a useful alternative when the H$\alpha$ region is unavailable (beyond z$\sim$0.4).  
    \item The Mass-Excitation (MEx) diagram and the Colour-Excitation (CEx) diagram provide more viable alternatives with the inclusion of stellar mass and (U-B) colour index respectively. In the present study of the GOODS-S field, the MEx diagram outperforms the other alternative diagnostics, e.g. correctly identifying a larger fraction of X-ray-selected AGNs compared to the CEx and [\ion{O}{II}] BPT methods. However, the results from both the MEx and CEx diagrams are highly correlated, and prove effective in identifying AGNs across a broader parameter space, especially at higher redshifts.  
    \item Cross-validation with X-ray-selected AGNs from the Chandra Deep Field South (CDFS) survey shows that optical diagnostics alone miss a significant fraction of AGNs ($\sim$65\%-70\%), reinforcing the necessity of multi-wavelength approaches.  
    \item Host galaxy age estimates indicate that AGNs preferentially reside in older host galaxies, supporting the scenario in which black hole growth becomes more dominant as galaxies evolve.  
    \item We have presented a methodology whereby the age as well as lookback time of a galaxy can be estimated from its optical spectrum. If utilised correctly, this provides an immensely valuable tool to understand the evolution of galaxies, and hence of the universe, on cosmic time scales.
    \item \autoref{fig_7} presents a novel approach to studying the co-evolution of star-forming galaxies and AGNs. By analyzing lookback time in combination with host galaxy age, we can directly probe how AGN activity and star formation have evolved over cosmic time. This temporal evolution analysis reveals that AGN-candidate formation peaks earlier ($\sim$8 Gyr ago), compared to the average SF-galaxy formation ($\sim$6 Gyr ago), although star formation undoubtedly remains widespread across all epochs.  
    \item  Furthermore, we can conclude that AGN activity peaked in an era of higher cosmic temperature ($T_\text{CMBR} = 5.8$ K at $\tau_\text{tot} \sim 5$ Gyr), likely driven by enhanced gas accretion.
    \item By combining lookback time with galaxy ages derived from spectral fitting, we can probe galaxy evolution all the way to early cosmic epochs (up to $\sim$13 Gyr ago), even though the observed sample is limited to galaxies at $z \sim 1$ (with lookback times of only $\sim$8 Gyr).
\end{enumerate}  

In conclusion, our findings emphasize the importance of integrating host galaxy modeling with multi-wavelength data to improve AGN classification. By extracting host galaxy age from optical spectra, we add an important — and hitherto overlooked — dimension to understanding galaxy evolution. While redshift indicates when we observe a galaxy, stellar age reveals how long it had been evolving before that point. Together with lookback time, this enables us to estimate when a galaxy likely formed, offering a more complete view of its life history and the physical conditions that shaped it. Our spectral fitting framework, combined with both traditional and extended AGN diagnostics, shows that AGNs are preferentially found in older galaxies and likely formed earlier in cosmic history than star-forming systems. This work demonstrates that even limited-redshift ($z \lesssim 1$) optical surveys can yield meaningful insights into galaxy evolution when host galaxy age and lookback time are jointly analyzed. The observed trends can serve to inform future studies on the role of physical conditions — such as cosmic background temperature — in shaping galaxy evolution.

\section*{Data Availability Statement}

This research made use of data obtained from the European Southern Observatory (ESO) Science Archive Facility, in accordance with the ESO Data Access Policy. Observations were originally carried out using the FORS1 and FORS2 instruments \citep{1998_Appenzellar}, and the VIMOS instrument \citep{2003_Le}. We made use of the publicly available FORS2 GOODS dataset (DOI: 10.18727/archive/28) and the VIMOS GOODS dataset (DOI: 10.18727/archive/31), obtained via the ESO Science Archive Facility. A complete list of ESO archival DOIs is available at \url{https://archive.eso.org/wdb/wdb/doi/collections/query}.

\section*{Acknowledgements}

This work also made extensive use of the Python programming language and several open-source scientific libraries, including NumPy \citep{2020_Harris}, SciPy \citep{2020_Virtanen}, Pandas \citep{2010_Mckinney}, Matplotlib \citep{2007_Hunter}, and Astropy, a community-developed core Python package for astronomy \citep{2022_Astropy}. We gratefully acknowledge the developers and maintainers of these packages for making their tools freely available to the scientific community.

GP acknowledges support provided by the Polish National Science Center (NCN) through grants no. 2023/49/B/ST9/01671, and 2024/53/N/ST9/03885.

The authors also thank Prof. M. N. Vahia, Dr. A. N. Raghav and Mr. B. Palit for their valuable comments and suggestions during the preparation of this work.




\bibliographystyle{mnras}
\bibliography{reference_list} 




\appendix

\section{Spectral Fitting Algorithm}

\begin{figure*}
    \centering
    \includegraphics[width=\textwidth]{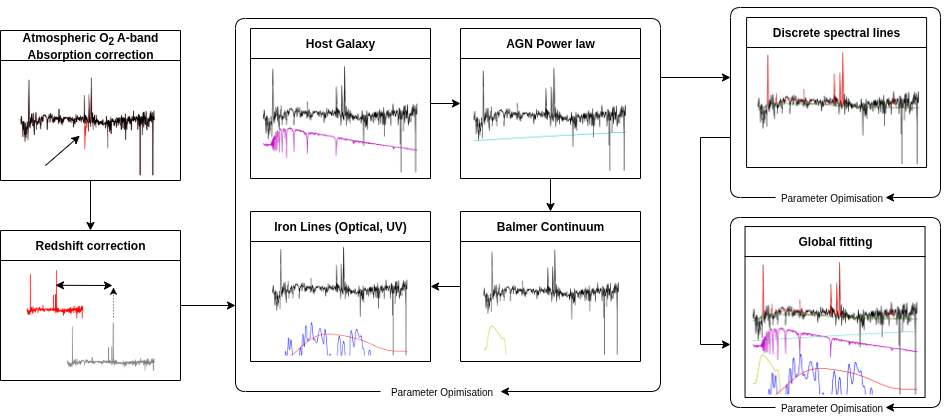}
    \caption{Flowchart illustrating the spectral fitting pipeline used in this work. The process begins with data preprocessing steps, including telluric correction and redshift shifting to the rest frame. The baseline fitting stage models the stellar continuum using BC03 templates, AGN emission with a broken power-law, Balmer continuum around the 3646\,\AA\ break, and Fe\,\textsc{ii} pseudo-continuum features in the optical and UV regions. Emission lines such as H$\alpha$, H$\beta$, [\ion{O}{III}]$\lambda$5007, [\ion{N}{II}]$\lambda$6584, and others are then fitted using Gaussian profiles with physically motivated constraints on width and velocity offset. A final global fit jointly refines both continuum and line parameters. The code outputs all derived quantities and fitted spectra, enabling detailed analysis of stellar populations, AGN activity, and gas properties.}
    \label{Fig_A1}
\end{figure*}

In this appendix, we describe the spectral fitting methodology adopted in this paper. The current status of the scientific literature in this field has been summarized in the main body of the paper (see \autoref{Fitting Section}). We focus here only on the methodology used.


The spectral fitting process in our code is primarily divided into two main stages: baseline fitting, and spectral peak fitting. In addition, certain data pre-processing is necessary, which constitutes a preliminary stage. The process begins with correction for atmospheric absorption features such as the telluric oxygen dip at 765 nm using methods described by \cite{2008_Tran_Hartmann_JGR}. The spectra are then redshift-corrected using redshift values from literature (see Sec. \ref{Data_section} and Table \ref{redshift_surveys}). Once the spectrum is shifted to the rest frame, we fit the continuum baseline. For this, the characteristic emission peaks are masked to prevent them from biasing the modeling of the underlying continuum.

In the baseline fitting stage, the continuum is modeled as a combination of multiple physical components, each contributing to the observed spectrum. The continuum model includes a broken power-law component (see e.g. \cite{2017_Calderone_MNRAS}) to represent the non-thermal AGN emission. This power-law is characterized by the luminosity density and spectral slope, capturing the broadband behavior of the AGN continuum. Previous studies have adopted varied strategies to model the host galaxy contribution in AGN spectra. For instance, \citet{2011_Shen_ApJS} employed a geometric mean of composite spectra, while \citet{2017_Calderone_MNRAS} modeled the host using a 5 Gyr elliptical galaxy template from \citet{1998_Silva}. More recently, \citet{2023_Ilic_ApJ_Fantasy} utilized a linear combination of normal galaxy and quasar eigenspectra. In our code, we have incorporated a state-of-the-art host galaxy stellar contribution model, using templates from the BC03 library \citep{2003_Bruzual_Charlot_MNRAS}, which includes a wide range of stellar population ages (0.1 Myr to 14 Gyr) and metallicities ($Z = 0.0001$ to $0.06$). During the fitting of these templates the component's parameters are varied until the differences between the data and the model are minimized, following a least squares minimization algorithm. The goodness of fit is assessed using the R-squared statistic. 

In addition, the baseline model in our code also includes the Balmer series continuum. The continuum above the Balmer series limit (3646~\AA), and the closely spaced discrete Balmer lines constituting a pseudo-continuum are modelled based on \cite{1985_Wills_APJ,1982_Grandi,2002_Dietrich}. Lines up to principal quantum number $n=40$ are included. 

Yet another (pseudo) continuum consists of broad and narrow Fe lines, which may especially overlap with the H$\beta$ region. We incorporate these \ion{Fe}{II} line templates. \ion{Fe}{II} lines in the UV region are taken from the \citet{2001_Vestergaard_Wilkes_APJS} template, while optical \ion{Fe}{II} features are from \citet{2004_Veron-Cetty_AandA}. These templates are convoluted with Gaussian profiles whose widths are constrained to physically meaningful ranges: 100--2000~km~s$^{-1}$ for narrow components and 1000--15{,}000~km~s$^{-1}$ for broad ones.

The construction of the baseline model proceeds by sequentially fitting and summing each component in a physically motivated order. First, the stellar continuum is modeled using the BC03 templates, establishing a baseline that reflects the host galaxy’s stellar light. Next, the AGN continuum is added via the broken power-law, capturing the broad-band non-thermal contribution. The Balmer continuum is then incorporated to account for the pseudo-continuum near 3646~\AA. Finally, \ion{Fe}{II} templates are fitted and added, fine-tuning the model particularly around the H$\beta$ and UV regions. Throughout this process, masking is applied to exclude prominent emission lines to avoid biasing the continuum fitting. This multi-step fitting ensures that each physical component is isolated and constrained before the full model is synthesized.

Once the baseline continuum has been fitted to the masked regions of the spectrum, the final stage involves fitting the full (unmasked) spectrum, including discrete emission lines. These are modeled within localized wavelength windows using Gaussian profiles. The emission lines include [\ion{O}{ii}], [\ion{Ne}{iii}], H$\delta$, H$\gamma$, H$\beta$, [\ion{O}{iii}], \ion{He}{i}, H$\alpha$, [\ion{N}{ii}], and [\ion{S}{ii}]. Each line is characterized by its amplitude, full-width-at-half-maximum (FWHM), and velocity offset. FWHM values are constrained between 100–2000 km s$^{-1}$ for narrow components and 1000–15,000 km s$^{-1}$ for broad components, while velocity offsets are limited to $\pm1000$ km s$^{-1}$ (narrow) and $\pm3000$ km s$^{-1}$ (broad). Subsequently, the continuum and emission-line parameters are jointly optimized in a global fitting procedure.

The code also tabulates all derived parameters and stores the fitted models and spectral plots, facilitating further analysis of the gas composition, density, and dynamics in galaxies and AGNs. Such a global fitting approach allows us to decompose complex spectra into their constituent physical components and extract key astrophysical insights.



\bsp	
\label{lastpage}
\end{document}